# Niels Bohr in North America (1923) and his Meeting with Robert Frost

Helge Kragh*

**Abstract**: During the autumn of 1923, Niels Bohr stayed for about three months in Canada and the United States where he gave a series of lectures on atomic and quantum theory. This paper provides details about his many meetings and lectures. It deals in particular with his stay at Amherst College, where he had a long conversation with the celebrated poet Robert Frost. Based on new archival sources, the paper includes a more fine-grained and better documented version of the episode than the one offered by literary critics and scholars. It concludes, perhaps somewhat disappointingly, that the Bohr-Frost meeting was of no great importance to either the physicist or the poet.

## 1. From Copenhagen over Liverpool to Toronto

Still at about 1920 Niels Bohr's theory of atomic structure dating from 1913 was not generally known and nor was the founder of the theory a household name. For example, in a semi-popular book published in 1917, two American physicists described J. J. Thomson's theory as well as Ernest Rutherford's modern "nucleus theory" and its development by "the Dutch physicist, Bohr" (Comstock and Troland 1917, p. 181). By and large, it was only after Bohr was awarded the Nobel Prize in 1922 that non-physicists became aware of him and his theory. By the mid-1920s Bohr's theory was relatively well known also by an educated lay audience. Several popular or semi-popular books had appeared in English, among them John W. N. Sullivan's *Atoms and Electrons* and Bertrand Russel's *The ABC of Atoms*, both published in 1923. They were followed the same year by *The Atom and the Bohr Theory of Its Structure*, written by the physicist Hendrik A. Kramers and the librarian Helge Holst and carrying a foreword by Rutherford (Kragh and Nielsen 2013).

When Bohr was awarded the Nobel Prize, he had never been outside Europe. That changed the following year when he gave a series of invited lectures in Canada and the United States. Curiously, his lecture tour has never been described in detail in the historical literature. According to the *New York Times*, it was largely due to Bohr that atomic structure caught the attention of the general public. On February 3, 1924 the newspaper wrote:

---

* Niels Bohr Institute, University of Copenhagen, 2100 Copenhagen, Denmark. E-mail: helge.kragh@nbi.ku.dk. This is an abridged and revised version of a longer paper submitted to *Perspectives on Science* in June 2026.

> The atom is getting to be a leading topic of conversation nowadays, even in circles where it had never been discussed before except in relation to persons or things having been "blown to atoms." Dr. Niels Bohr … is responsible largely for this addition to popular conversation. Since he came to this country last fall to lecture on his theory of the structure of the atom at Yale University and elsewhere, there have been a remarkable display of interest in his discoveries.[1]

Bohr started his extended journey by first going to Liverpool, where he attended the annual meeting of the British Association for the Advancement of Science taking place 12-19 September with Rutherford as president.[2] On the 22nd he left England from Southampton on board the ocean liner "Aquitania" together with John McLennan, a Canadian physicist who was a former student of J. J. Thomson and now served as director of the physics laboratory of the University of Toronto. The two proceeded by railway from New York to Toronto, where Bohr gave three lectures and was shown McLennan's "magnificent laboratory" which recently had been provided with advanced low-temperature equipment to liquefy helium. It was with this equipment that McLennan and his student Gordon Shrum two years later solved the long-time puzzle of the green spectral line in the aurora borealis (Kragh 2009).

"Professor Neils [*sic*] Bohr, Great Theoretical Physicist Will Lecture Here." This was how *The Varsity*, a Toronto undergraduate newspaper, announced Bohr's lectures, adding that they would be "of a very technical nature to research students in Mathematical physics." The writer of the column referred to Bohr's successful theory of line spectra but without understanding what it was about: "These lines are caused by the vibrations of electrons which vibrate at different rates and the success of which he has arrived can be gauged from the fact that there are several hundred different kinds of electrons in the element iron."[3] Apart from delivering the lectures "of a very technical nature" Bohr also gave a talk to the students in which he reflected on the different conditions of science students in Copenhagen and Toronto. According to *The Varsity* of 3 October, Bohr recommended greater international exchange of students from different countries:

> Professor Bohr emphasized the great need of the present day as that of a better understanding between students throughout the world … He believes that students should spend some time in countries other than their own. In this way by meeting their fellow students in other countries and coming in contact with their difficulties they would learn not to look down upon one another but to grow in understanding.

---

[1] On January 20, 1924 the same newspaper stated: "One of the biggest snowballs in the new science is the theory of atomic structure advanced by Professor Niels Bohr of the University of Copenhagen."

[2] For Bohr's series of American lectures, see Nielsen 1976, pp. 45-46; Pais 1991, pp. 253-254; Pais 1989. However, these sources contain no description of the Simpson lectures and no mention of Frost. The present account is partly based on the letters that Bohr sent to his wife Margrethe during his journey. The correspondence between the two is kept at the Niels Bohr Archive (NBA) in Copenhagen, and is here cited with the kind permission of Vilhelm Bohr, a grandson of Niels Bohr.

The *Star Weekly*, a Toronto Magazine, covered Bohr's theory over a full page including some of his lecture slides. "Through a maze of mathematics and the computation of forces applied to those almost infinitely minute orbits … he developed in detail these various and surprisingly different solar systems, not forgetting the comet, or, rather, comet-like orbits, of some of the electrons."[4]

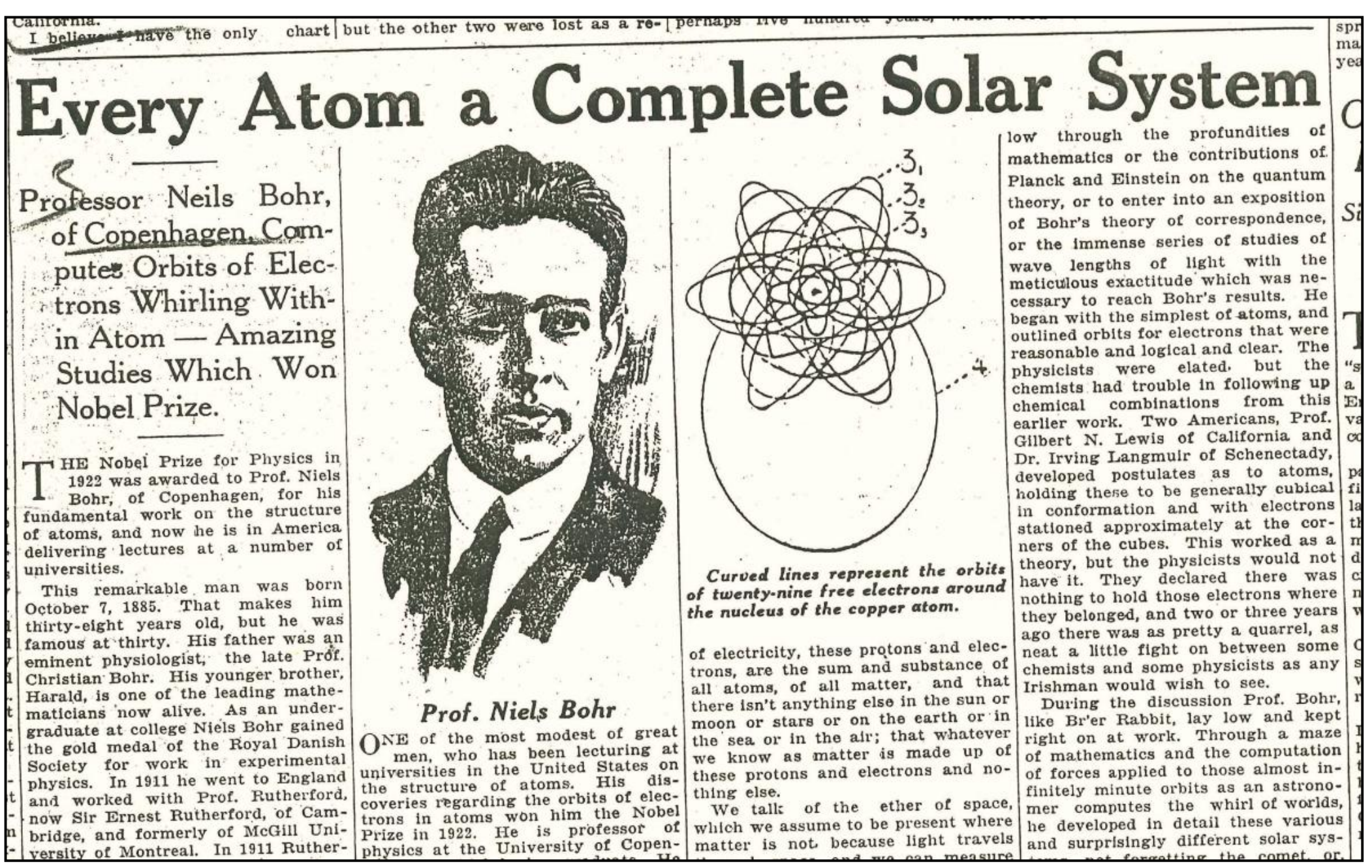

# Every Atom a Complete Solar System

**Professor Neils Bohr, of Copenhagen, Computes Orbits of Electrons Whirling Within Atom — Amazing Studies Which Won Nobel Prize.**

THE Nobel Prize for Physics in 1922 was awarded to Prof. Niels Bohr, of Copenhagen, for his fundamental work on the structure of atoms, and now he is in America delivering lectures at a number of universities.

This remarkable man was born October 7, 1885. That makes him thirty-eight years old, but he was famous at thirty. His father was an eminent physiologist, the late Prof. Christian Bohr. His younger brother, Harald, is one of the leading mathematicians now alive. As an undergraduate at college Niels Bohr gained the gold medal of the Royal Danish Society for work in experimental physics. In 1911 he went to England and worked with Prof. Rutherford, now Sir Ernest Rutherford, of Cambridge, and formerly of McGill University of Montreal. In 1911 Ruther-

*Prof. Niels Bohr*

ONE of the most modest of great men, who has been lecturing at universities in the United States on the structure of atoms. His discoveries regarding the orbits of electrons in atoms won him the Nobel Prize in 1922. He is professor of physics at the University of Copen-

*Curved lines represent the orbits of twenty-nine free electrons around the nucleus of the copper atom.*

of electricity, these protons and electrons, are the sum and substance of all atoms, of all matter, and that there isn't anything else in the sun or moon or stars or on the earth or in the sea or in the air; that whatever we know as matter is made up of these protons and electrons and nothing else.

We talk of the ether of space, which we assume to be present where matter is not because light travels low through the profundities of mathematics or the contributions of Planck and Einstein on the quantum theory, or to enter into an exposition of Bohr's theory of correspondence, or the immense series of studies of wave lengths of light with the meticulous exactitude which was necessary to reach Bohr's results. He began with the simplest of atoms, and outlined orbits for electrons that were reasonable and logical and clear. The physicists were elated but the chemists had trouble in following up chemical combinations from this earlier work. Two Americans, Prof. Gilbert N. Lewis of California and Dr. Irving Langmuir of Schenectady, developed postulates as to atoms, holding these to be generally cubical in conformation and with electrons stationed approximately at the corners of the cubes. This worked as a theory, but the physicists would not have it. They declared there was nothing to hold those electrons where they belonged, and two or three years ago there was as pretty a quarrel, as neat a little fight on between some chemists and some physicists as any Irishman would wish to see.

During the discussion Prof. Bohr, like Br'er Rabbit, lay low and kept right on at work. Through a maze of mathematics and the computation of forces applied to those almost infinitely minute orbits as an astronomer computes the whirl of worlds, he developed in detail these various and surprisingly different solar sys-

Fig. 1. The *Star Weekly* article on Bohr's atomic model.

[3] *The Varsity: The Undergraduate Newspaper* 43 (27 September, 1923). In Bohr's theory the emission of light was not caused by vibrating electrons but by sudden quantum transitions governed by the energy difference between two stationary states.

[4] "Every Atom a Complete Solar System," *Star Weekly*, December 8, 1923.

## 2. A busy travel itinerary

As Bohr reported to his wife Margrethe in a letter of October 7, "on Friday morning I left Toronto and saw the Niaga [Niagara] Falls on the border between Canada and the United States." Bohr arrived in Amherst, Massachusetts, on October 6, where he stayed for about three weeks. While in Amherst, he also visited Harvard University in Boston and the General Electric laboratory in Schenectady. His lecture in Harvard was briefly described in the October 26 issue of *The Harvard Crimson*, a daily college newspaper, under the headline "Professor Bohr Tells How He Invented the Model Atom":

> Over 300 people crowded the large room of Jefferson Physical Laboratory yesterday afternoon to hear Professor Niels Bohr of the University of Copenhagen give the first of his lectures on the theory of spectra and atomic constitution. Professor Bohr, who has won international reputation in connection with the theory of atomic constitution, in his lecture yesterday explained the model atom which he has invented.

Among those he met in Harvard was the noted physicist Percy Bridgman who in 1946 would receive the Nobel Prize for his work on matter under high pressure. He also met the chemistry professor Theodore Richards, with whom he discussed the still controversial status of element 72 in the periodic table. "I talked a great deal with him [Richards] concerning the hafnium case; he understood everything and was much enthusiastic about the work of Hevesy and Coster."[5] Guided by Bohr's atomic theory, in early 1923 his collaborators George Hevesy and Dirk Coster had discovered the new element, but the discovery claim was contested by French scientists who argued that priority belonged to them. Worried about the controversy, Bohr took up the subject in some of his American lectures.

All the way from Liverpool Bohr was accompanied and assisted by Frank Hoyt, a young American physicist who had stayed at the Copenhagen institute since 1922.[6] Bohr and Hoyt departed in late November after which Bohr continued his busy itinerary alone, visiting also Pittsburgh, Chicago, Ann Arbor, Rochester, Baltimore, and Princeton before he returned directly from New York to Copenhagen in mid-December on the ocean liner "Frederik VIII." Several American universities and research institutions on the West Coast also wanted Bohr to come and give lectures, but he regretfully declined the offers.

---

[5] Niels Bohr to Margrethe Bohr, October 28, 1923 (NBA). Theodore Richards (1868-1928) received as the first American the chemistry Nobel Prize in 1914. On the hafnium controversy, see Kragh 1980.

[6] Frank Clark Hoyt (1898-1980) was based in Copenhagen from the autumn of 1922 to the spring of 1924 working in particular with the mathematical formulation of the correspondence principle. In an interview of April 29, 1964 conducted by Thomas Kuhn he recalled his and Bohr's travels in 1923. See https://repository.aip.org/node/5540

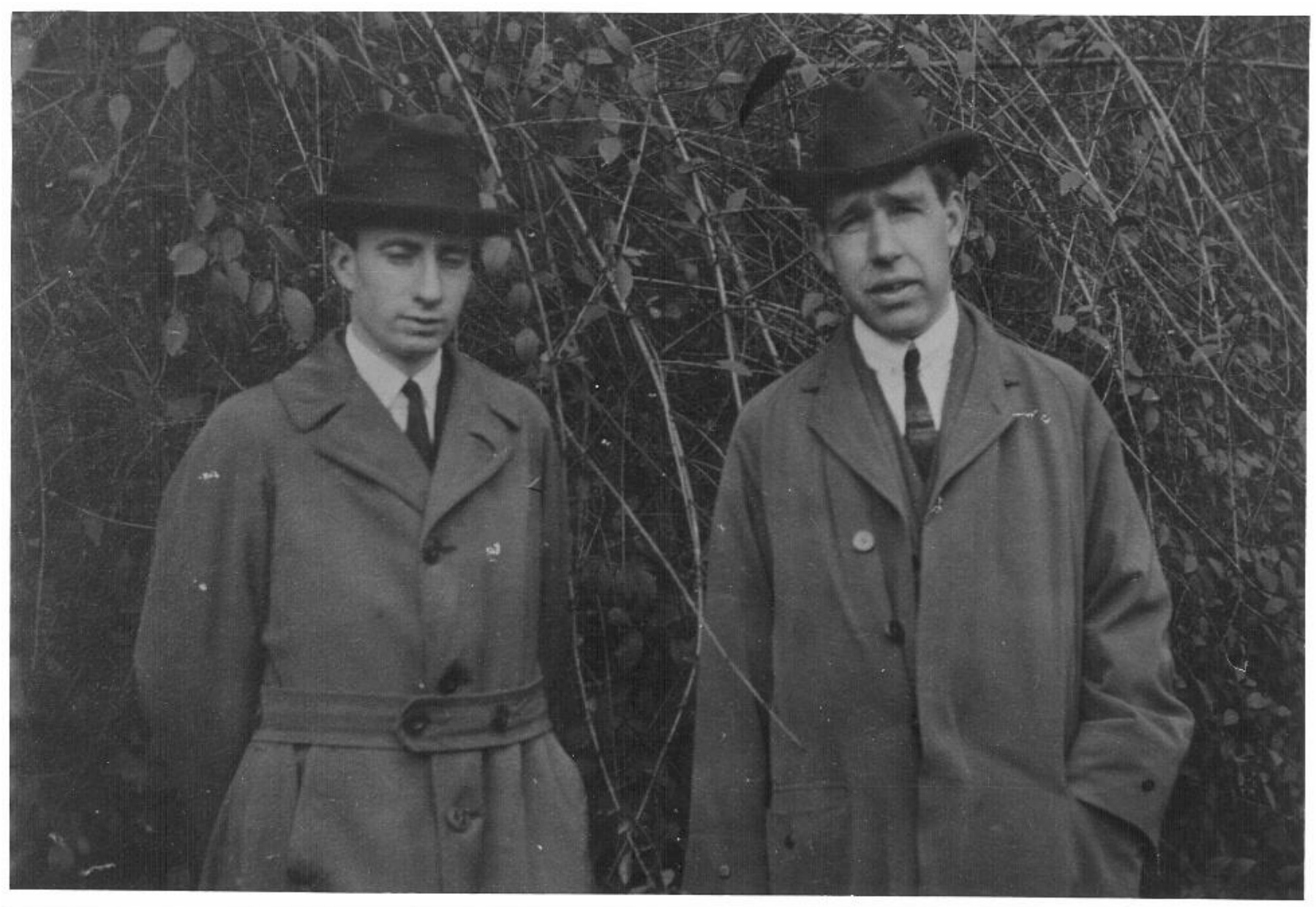

Fig. 2. Frank Hoyt and Niels Bohr, Copenhagen 1922. Courtesy NBA.

Bohr did not originally connect his travels in America with the possibility of fund raising. However, when he became aware of the recently established International Education Board (IEB) he used the occasion to apply for means to expand his institute in Copenhagen. Right after the first Silliman lecture, he had a meeting in New York with IEB officers such as he reported to Margrethe: "I was in New York this morning to meet President Rose (International Education Board). I had such a good meeting with him and I definitely believe that we will get the money we need."[7] Bohr was right. In late January 1924 the *New York Times* announced that he had received from the IEB the substantial sum of $40,000 "so that he can continue his revolutionary research into the structure of the atom."[8]

While still in Amherst, Bohr gave a talk on "Some Philosophical Aspects of Modern Work in Physics" which was attended not only by a local audience but also by a group of physicists from Schenectady.[9] Among them was the prominent chemist and later Nobel laureate Irving Langmuir, who had his own ideas concerning atomic structure. Langmuir and Bohr had corresponded about atomic theory in 1921 and now, two years later, they met face-to-face. "I am looking forward very much indeed to meeting you in Amherst on Monday, Oct. 29th," Bohr wrote. "It shall be a great pleasure to me to accept the kind invitation of the General Electric and go back with you to Schenectady on Tuesday Oct. 30 to stay for a few days."[10] To Margrethe he told enthusiastically about the motor tour through the scenic countryside and the huge General Electric works, "the most impressive I have ever seen." He concluded: "It was a great and wonderful experience in Schenectady, but the greatest joy was however to learn about Langmuir as a person; he is a highly intelligent and charming man and I now think we know each other much better."[11] Langmuir reciprocated his sentiments and that to an even higher degree. According to Langmuir's biographer, "Over the years Langmuir's respect and affection for Bohr grew into something close to hero worship" (Rosenfeld 1966, p. 188).

[7] Niels Bohr to Margrethe Bohr, November 5, 1923 (NBA). See also Pais 1989. Wickliffe Rose (1862-1931) directed the IEB from 1923 to 1928.

[8] *New York Times*, January 28, 1924.

When visiting Ann Arbor, Bohr met with his former assistant and close collaborator Oskar Klein, who had recently taken up a position as instructor at the University of Michigan. Klein only returned to Copenhagen in early 1926 and then stayed at Bohr's institute until he moved permanently to Stockholm in 1931. In Ann Arbor the Swedish physicist worked with David Dennison, then a 23-year-old graduate student, who later recalled: "He [Bohr] came no doubt because Klein had been his student, he came here and gave a lecture. And Bohr was very much interested in what I was doing and talked with me a good deal."[12] In fact, it was on Bohr's instigation that Dennison in October 1924 went to Copenhagen where he soon became introduced to the new quantum mechanics and did important work in atomic and molecular physics. He returned to Michigan in 1927. Thus, Bohr's brief stay in Ann Arbor was not without consequences.

On November 24, about two weeks before his return, Bohr attended a colloquium at the National Bureau of Standards in Washington D.C. which was followed by a dinner in his honour. According to *Science*, Bohr "spoke briefly on the great possibilities just ahead in the field of science, likening the present to the time of Newton which preceded great things in the scientific world."[13] Other speakers at the dinner party included Paul D. Foote, a specialist in spectroscopy, Charles G. Abbot, a noted astrophysicist, and Fredrick G. Cottrell, a physical chemist and inventor. Cottrell suggested somewhat cryptically that Bohr's latest theories might lead to "real progress in the fixation of nitrogen."

---

[9] The contents of the lecture are indicated by notes taken by Hoyt, see Nielsen 1976, pp. 45-46. One of the subjects that Bohr spoke of was "Questionable character of gain by developing picture of classical ideas" and another was "Analysis of what explanation means in science."

[10] Bohr to Langmuir, October 29, 1923 (NBA)

[11] Niels Bohr to Margrethe Bohr, November 4, 1923 (NBA). Bohr had earlier criticized Langmuir's chemical atomic theory in strong terms, see Kragh 2012, pp. 231-235.

[12] Interview by Thomas Kuhn of January 28, 1964. https://repository.aip.org/node/127863

Three days after the Washington meeting, Bohr reported in one of his letters to his wife: "From New York I went to Baltimore, where I spent a very interesting day together with Wood; you know, the famous physicist who masters everything, who paints and writes children's books, and so on." He referred to Robert Williams Wood, a professor at Johns Hopkins University and a renowned specialist in infrared and ultraviolet spectroscopy. His work in this area had been important to Bohr's 1913 quantum theory of atomic structure and Bohr valued it greatly. As indicated in the letter, Wood also wrote fiction novels and light-hearted poetry such as the charming *How to Tell the Birds from the Flowers* from 1907. Incidentally, in one of Wood's science fiction novels, *The Man Who Rocked the Earth* co-authored by Arthur Train, there appears a brief reference to quantum theory, possibly the first in fiction literature.[14] Bohr and Wood had both attended the British Association meeting in Liverpool, where they had discussions about atoms and radiation. When in Baltimore in November, they continued their discussions.[15]

## 3. The Silliman lectures

In October 1923 Bohr gave the five Simpson lectures at Amherst College, an undergraduate institution founded in 1824, where he dealt with the structure of atoms, radioactivity, the correspondence principle, the nature of light, the periodic table, and other topics of modern physics. These were the same topics that he covered in his better known and more prestigious Silliman lectures at Yale the following month, only were they presented in a less technical manner in Amherst. Bohr was supposed to prepare a book manuscript on the basis on his lectures in Yale, such as previous Silliman lecturers had done. Thus, his compatriot August Krogh, a physiologist and Nobel Prize winner of 1920, had delivered the Silliman lectures for 1922 and his book appeared promptly the same year (Krogh 1922). Bohr's planned book never materialized.

---

[13] "Dinner in honor of Dr. Bohr," *Science* **58** (December 28, 1923): 533-534.

[14] Train and Wood 1915. Between 1914 and 1950 Wood was nominated for the Nobel Prize no less than 39 times, twice with Bohr as a nominator. Pais 1991, p. 216.

[15] Wood to Bohr, October 25, 1923 and Bohr to Wood, November 1923 (NBA).

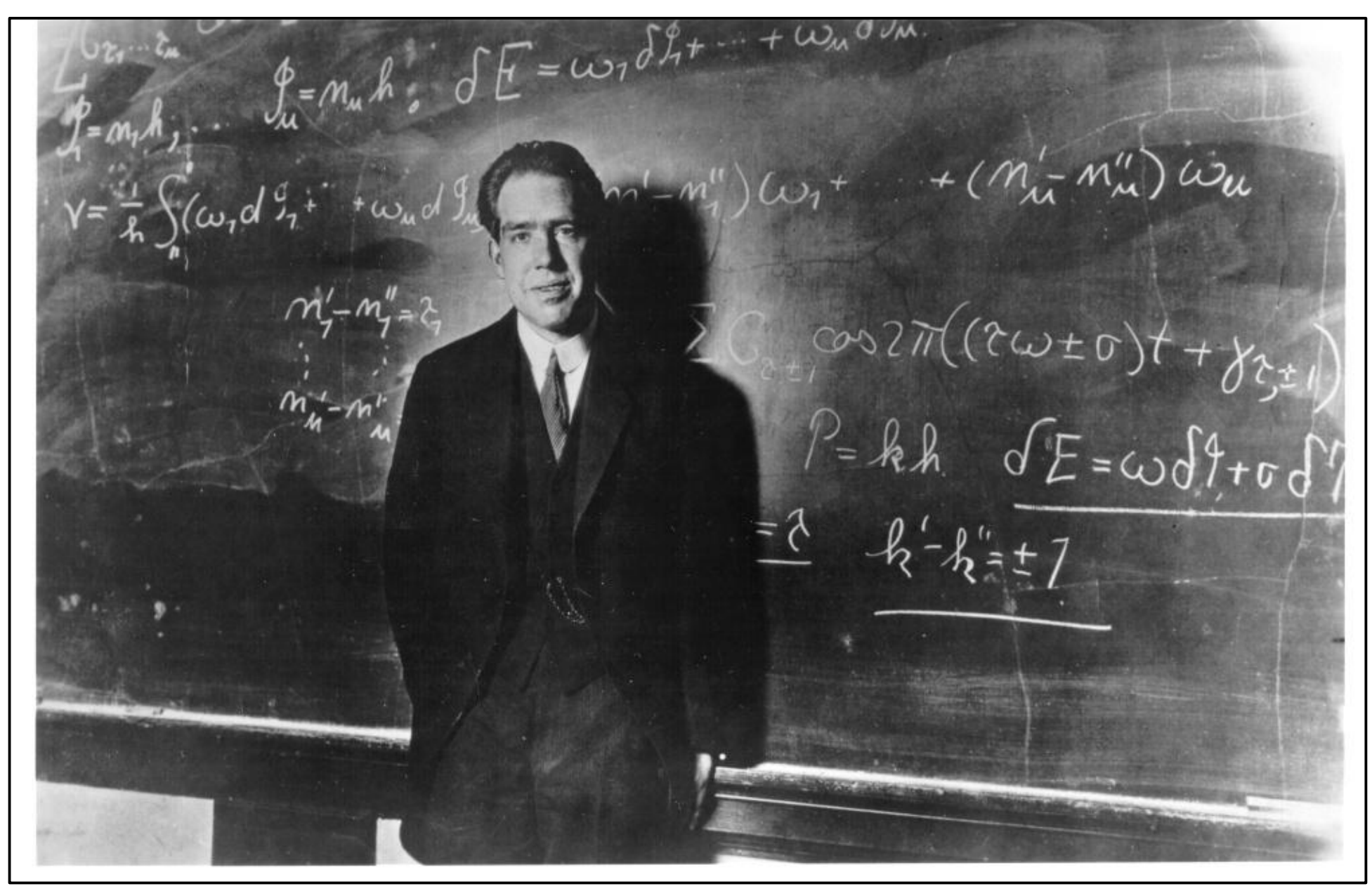

Fig. 3. Bohr giving one of his Silliman lectures. Courtesy NBA.

As noted in *Science* of 19 October: "In addition to giving the Silliman lectures at Yale University, Dr. Niels Bohr, professor of physics in the University of Copenhagen, is giving the Simpson lectures at Amherst College." While American national newspapers bypassed the lectures in Amherst, those delivered at Yale made headlines. *New York Times* covered the Silliman lectures in considerable detail in an article of November 7 headlined:

> DR. BOHR EXPOUNDS THEORY OF ATOMS: Begins a Series of Lectures at Yale on His Study of Revolving Electrons. LIKENED TO SOLAR SYSTEM: He Pictures the Atom with Nucleus Corresponding to Sun, and Electrons to Planets.

Reporting from the first lecture on November 6, the newspaper informed its readers that "The hall was crowded this afternoon with students, professors and those of the general public specially interested in the subject. Experts from electrical companies from New York and elsewhere were present." The reporter cited from the lecture:

> We assumed twenty years ago that we might be able to work down some day to a picture of the atom from larger bodies of matter; but now we have been presented with a very definite picture of the atom from other sources. All doubt regarding the existence of the atom has disappeared … We now have a picture of the atom which we

> believe I think all physicists now believe is just as real as any of the natural phenomena that we are in the habit of discussing.

Bohr also stressed that the new picture of the atom was after all very different from the popular solar system analogy:

> He pointed out that, although the atom was similar to the solar system in some respects, it was dissimilar in others; therefore, the ordinary laws of mechanics and electro-dynamics could not be used in the study of the atom ... If the ordinary laws held good for the atom, he declared, electrons would fall into the nucleus "and destroy the life of the atom," very much as if the earth or some other planet fell into the sun and destroyed the life of the universe.

The *New York Times* article ended with quoting John Zeleny, chairman of the Physics Department at Yale University: "We are fortunate in this course of Silliman lectures to have the new ideas presented to us by their great apostle."

From an article the following day, reporting on the second lecture, it appears that Bohr critically discussed Einstein's light quantum hypothesis. Although he recognized its explanation of the photoelectric effect, he did not believe that light consisted of particles. "The theory of light quanta is not to take the place of the old [wave] idea of light," Bohr asserted. At the time he was well aware of Arthur Compton's classic scattering experiments with X-rays which were widely seen as strong support of light quanta. However, Bohr disagreed.

During his visit to the United States he discussed the question with, among others, the X-ray specialist William Duane who confirmed his suspicion that the Compton effect could be explained on the wave theory of light. "I am still inclined to believe that the explanation I suggested during our conversation at Harvard may be correct," Bohr wrote from Copenhagen.[16] He also had a meeting on the subject with Compton and the physicist William Swann in November 1923, possibly in Chicago, where he attended a meeting of the American Physical Society giving a keynote address on "The Quantum Theory of Atoms with Several Electrons" (Pais 1991, p. 253; Kragh 2012, pp. 326-327). Bohr returned to Copenhagen confident that light quanta did not belong to the real world. While in Chicago he also met the famous Albert Michelson, America's first science Nobel Prize laureate, "who I believe found in me a more conservative scientist than he had expected."[17]

---

[16] Bohr to Duane, February 6, 1924, in Stolzenburg 1984, p. 320.

[17] Bohr to Rutherford, January 9, 1924, in Stolzenburg 1984, p. 487; Bohr to Michelson, February 7, 1924, same source pp. 404-405.

## 4. Bohr in Amherst: The Simpson lectures

The five Simpson lectures given on October 12, 16, 18, 22, and 24 were of a general nature and followed by some of the same slides that Bohr had used in his Nobel lecture in Stockholm. Amherst College had no research program in physics and so, as the president Alexander Meiklejohn stated, Bohr's lectures were "intended particularly to meet the needs of teachers in secondary schools" (Kojevnikov 2020, p. 50). As Bohr explained to his wife, he needed to present his lectures in a more popular manner: "There are no one here with a real knowledge of physics, but there is some general interest and I will strive to the best of my ability to make everything as simple as possible."[18] Although Bohr's visit to Amherst was of no scientific importance, it is of interest for other reasons.

In his notes for the first lecture Bohr stated that the purpose was "to show that the starting point has been in experimental discoveries and not in speculations." His notes for the final fifth lecture included a reference to existing atomic theory's failure in explaining the covalent bond and the difference between this theory and the one favoured by the chemists: "Polar molecules [ionic compounds] depending only on constitution of ions. Non-polar molecules essentially more complex. No argument however by qualitative chemical considerations gives a theory of orbits which is essentially quantifiable in character." At the end of his lecture notes, he wrote: "Reality in the sense that we deal with unambiguous connection between experimental facts. Incomplete as regards picture. New conceptions of natural philosophy. Reasons for fascination of theory. Hope to have created interest."[19] In his more extensive notes for the Silliman lectures, Bohr addressed the same issues. Thus, he mentioned the "difficulties in accounting for homeopolar [covalent] compounds" and concluded by emphasizing "the reality of the theory as well as its incompleteness."[20]

The contents of the four Simpson lectures were reported in the local *Amherst Student*, from which some further information can be gained. About the last lecture, which was "well attended by faculty, students and townspeople" the student newspaper wrote: "Slides were used to discuss the particularly fine structure of hydrogen lines, the spectro-diagram of the Stark and Seeman [Zeeman] effects being the most important part of the lecture. Professor Bohr also gave an elementary exposition of the correspondence principle, which was followed by a discussion of the quantum theory to systems of several electrons" (quoted in Ahearn 2020, p. 164).

---

[18] Niels Bohr to Margrethe Bohr, October 14, 1923 (NBA).

[19] "Amherst." Handwritten notes from 10 to 22 October 1923 (NBA).

[20] Nielsen 1976, p. 601. The contents of Bohr's Silliman lectures were summarized in *Science* 58 (December 7, 1923): 459-460.

As mentioned, most of the popular expositions of the Bohr atom employed the planetary analogy describing the nucleus (the Sun) as a large body surrounded by the much smaller electrons (the planets). As pointed out in a detailed and informative article in *Popular Radio*, the metaphor was unfortunate given that in the Rutherford-Bohr model the electron was actually very large compared to the nucleus. On the widely shared assumption that the mass of an elementary charge is of electromagnetic origin it follows that mass $m$ and radius $a$ varies as $m \sim a^{-1}$, that is, the greater the mass the smaller the volume. *Popular Radio* asked its readers to imagine a hydrogen atom at the size of the orbit of the Earth: "In our enlarged hydrogen atom where the electron is 6,900 miles in diameter, the central proton will have a diameter of *less than four miles*" (Free 1924; summary in *The Literary Digest*, April 12, 1924). Bohr never spoke of the size of electrons.

## 5. Robert Frost on science

Forty-six-year-old Robert Frost, who the same year had published the poetry collection *New Hampshire* (for which he was awarded a Pulitzer Prize in 1924) taught at the time a course in English literature at Amherst. Throughout his life he had close connections to the college, where he for longer periods served as professor of English literature. Frost was interested in science generally, and in astronomy in particular, and followed its development through *Scientific American* and other similar magazines.[21] But he was also critical, having no faith in science as a royal road to truths about nature. "Science has half the world, the lower half," he said. "All science is domestic science – domesticated on earth." There were different kinds of truth and Frost rated poetic truth higher than scientific truth. Concerning the relationship between science and poetry, he stated: "Science cannot be scientific about poetry, but poetry can be poetical about sciences. It's bigger, more inclusive."

It is not known whether the poet actually listened to any of Bohr's lectures, but it is known that on the occasion he had a conversation with the physicist. This is what he said in a later address, "Education by Poetry" first published in the February 1931 issue of *Amherst Graduates' Quarterly*. Frost was intensely interested in metaphors and tended to believe that they were no less important in mathematics and physics than they were in poetry and daily life. Science, he claimed, was essentially metaphoric: ''Poets and scientists have in common the biggest thing of all

---

[21] On Frost and science, see Smith 2018 and Abel 1980. The two following quotations are from the latter source.

– their metaphors. The poet and the scientist think by metaphor" (Mertins 1965, p. 368). He boiled down relativity theory to just a metaphoric explanation: "It seems to me that it is simply and utterly charming – to say that space is something like curved in the neighborhood of matter."

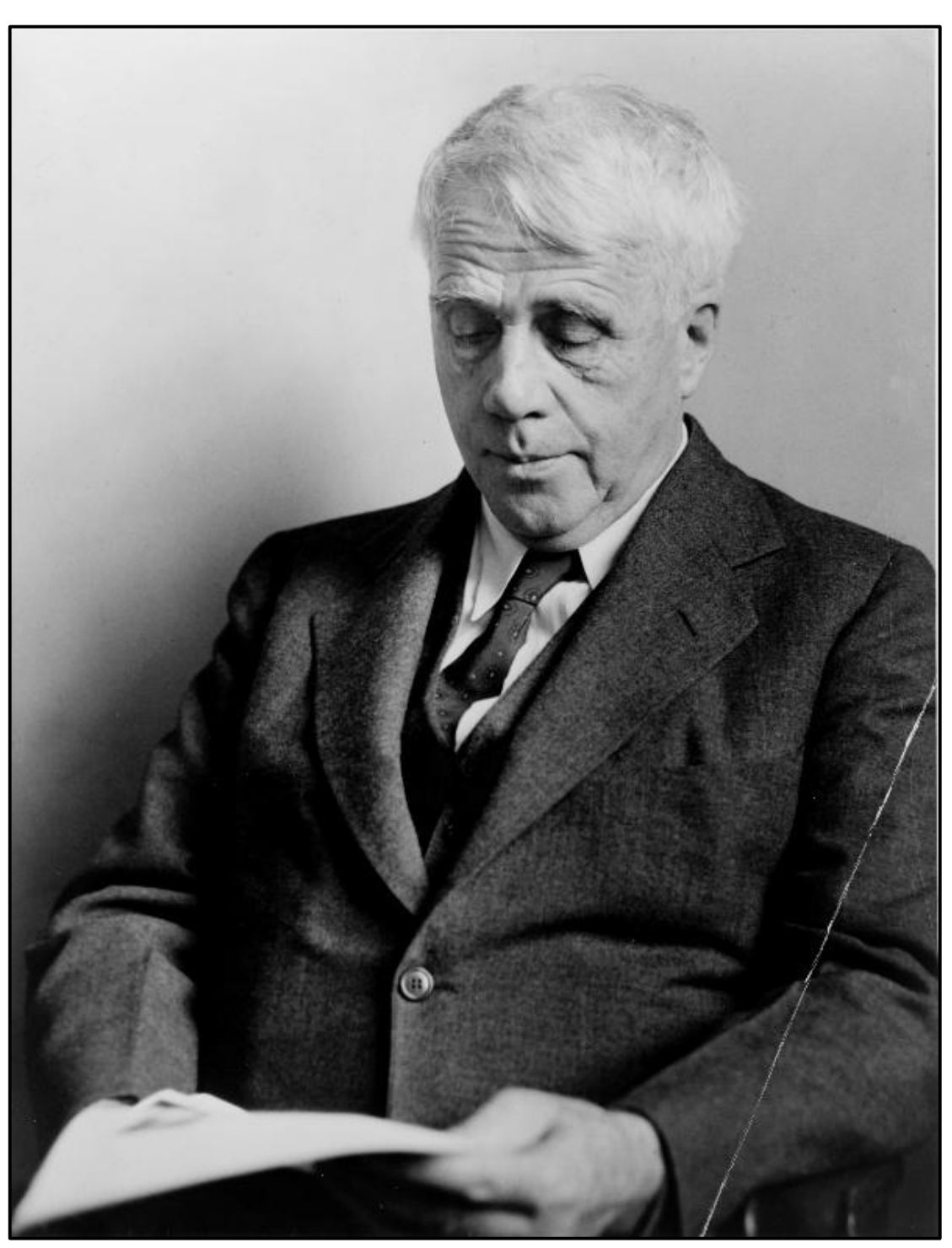

Fig. 4. Robert Frost in about 1940. Wikimedia Commons.

As another example of a charming scientific metaphor, Frost mentioned Heisenberg's position-momentum uncertainty relation $\Delta p \Delta q \geq h/2\pi$. According to the poet: "The other day we had a visitor here, a noted scientist, whose latest word to the world has been that the more accurately you know where a thing is, the less accurately you are able to state how fast it is moving … the more accurately you state where a thing is, the less accurately you will be able to tell how fast it is moving." Frost did not give the name of the noted scientist, but he probably had in mind Heisenberg, who in the spring of 1929 stayed as a visitor at the University of Chicago and the following year, according to one source, visited Amherst College.[22] 'Freedom' was one of Frost's favourite metaphors (Frost 1931; Ahearn 2020, p. 163):

---

[22] Smith 2018, p. 201 states that Frost "had the opportunity to meet Heisenberg, who visited Amherst College in 1930." However, there is no documentation known to me that he actually met Heisenberg or that Heisenberg was in Amherst in 1930. As suggested in Rotella

> You know that you can't tell by name what persons in a certain class will be dead ten years after graduation, but you can tell actuarially how many will be dead. Now, just so this scientist says of the particles of matter flying at a screen, striking a screen; you can't tell what individual particles will collide, but you can say in general that a certain number will strike in a given time. It shows, you see, that the individual particle can come freely. I asked Bohr about that particularly, and he said, "Yes, it is so. It can come when it wills and as it wills; and the action of the individual particle is unpredictable. But it is not so of the action of the mass. There you can predict." He says, "That gives the individual atom its freedom, but the mass its necessity."

The apparent freedom of electrons was not a new theme as it had been raised by sceptics in the early days of the Bohr atom. There is no strict rule or law that orders an atom in an excited state $n_2$ to decay to a particular lower state $n_1$ rather than some other state. The electron apparently decides in advance where to move, as had it its own free will. "It may be," wrote the British physicist Charles G. Darwin to Bohr in 1919, with tongue in cheek, "that it will prove necessary … in the last resort to endow electrons with free will" (Kragh 2011). A *New York Times* article used the 'freedom' of orbital electrons to illustrate how radically Bohr's theory departed from established natural philosophy. The theory "assumes that an action can depend upon the future as well as the past – that is, that an electron must 'know' where it is going before it can 'decide' how much energy to emit in its journey."[23]

## 6. The Frost-Bohr conversation and its literary aftermath

According to Frost's biographer Lawrance Thompson, the poet and the physicist had dinner "in the home of president Olds, who subsequently said that RF's questions addressed to Bohr were more penetrating than those asked by the professional scientists in the dinner group."[24] Another source states: "Frost was in the audience. The poet developed a profound curiosity about this new science of quantum mechanics and physics. President Olds invited Bohr and Frost to his home for dinner, and the two talked for hours" (D'Arienzo 2008).

In a letter of 1941 to his grandson Prescott, Frost summarized the then obsolete Bohr atom: "Niels Boar [*sic*] had the idea that the particles within the atom were

---

1987, the "noted scientist" may have been "a figment, a rhetorical device for creating authority and immediacy."

[23] R. Porter, "Pictures the Atom as a Small Solar System," *New York Times*, January 6, 1924.

[24] Thompson 1971, p. 617, citing an interview of 1947 with Mrs. Olds. RF = Robert Frost. George D. Olds (1853-1931), a former professor of mathematics, served as president of Amherst College from 1924 to 1927 and was acting president at the time of Bohr's lectures.

arranged and behaved like a minute solar system" (Smith 2018, p. 244). Nearly forty years after the event in Amherst, Frost recalled: "I've been close to men like Niels Bohr. Had great talks with him. Words and science come close together … No conflict."[25] According to Reginald Cook (1974), an English professor, "Frost greatly admired Bohr … [and] never tired of referring to Bohr's … great insights in modern science with the metaphor of planetary motion." And yet it is difficult to find in Frost's poems such references, or just allusions, to Bohr's atomic model described by one physicist as "a piece of scientific poetry of the highest art" (Saunders 1924).

Referring to the 1923 Frost-Bohr meeting, Robert Hass admits that "What the two talked about is open to speculation." One of the speculations, presented as were it a fact, is that the second law of thermodynamics was "one of the subjects RF discussed during his encounter with Niels Bohr" (Richardson, Hass and Atmore 2017, p. 508). According to Hass, the poet and the physicist also discussed how the new physics affected the old idea of a free will. Moreover: "Given that Bohr himself had read extensively from William James, Frost may even have prodded the physicist on several Jamesian issues" (Hass 2002, p. 109). This is possible, but the premise is open to doubt. Although Bohr knew about the philosopher and psychologist William James at the time, he had not read him extensively. His knowledge of and interest in James' *Principles of Psychology* seems to have been of a later date, after he had introduced the complementarity principle (Favrholdt 1992, pp. 63-73). The mentioned suggestions have in common that they are purely speculative.

Without referring specifically to the Amherst conversation, Rotella (1987) argues, such as Cook did, that Frost and Bohr had much in common. "The two might have found common ground in a discussion of language, ideas about which were fundamental to Bohr's thinking." Indeed, Bohr later became fascinated by language as a necessary resource not only of quantum physics but of all rational communication. His view is commonly encapsulated in the often quoted "We are suspended in language" phrase. He was convinced that the essence of scientific knowledge is that it allows unambiguous communication in terms of words. Literary scholars also like to cite Bohr for the statement that "When it comes to atoms, language can be used only as in poetry" which presumably would have appealed to Frost. However, all this is irrelevant with regard to the 1923 conversation. Besides, none of the two quotations are from Bohr. The first one was reported by his assistant

[25] Rotella 1987. The recollection is not literally Frost's, but as recalled by Louis Mertins in a conversation he had with Frost in the early 1960s and which he reported in Mertins 1965, p. 399.

Aage Petersen in 1963, after the death of Bohr, and the second one is due to Heisenberg's somewhat doubtful recollections from 1971 of his conversations with the Danish quantum sage.[26]

From Bohr's correspondence with his wife Margrethe we get a more fine-grained and authentic picture of Bohr's stay in Amherst and his meeting with Frost. On the 14th of October Bohr and Hoyt were dining with president Olds, whom Bohr described as "the most lovable old man you can think of." Ten days later, after having completed the Simpson lectures, Bohr again spent an evening at Olds' residence, this time together with Frost and Frederick Woodbridge, a philosopher and Dean at Columbia University.

> He [Woodbridge] was a most interesting man, and I talked with him for long and in detail about Kierkegaard and other philosophical subjects. There was also a Professor Frost here from Amherst who is a significant poet; in fact, some consider him to be the most significant of all Americans still alive. Unfortunately, I have not yet found time to read some of his books and this evening I did not talk with him at all.[27]

However, Bohr got to know Frost the next day when Olds invited the two and also Hoyt for an evening conversation. In his letter to Margrethe:

> Professor Old [*sic*], who is such a nice man, invited Hoyt and me to meet him also on Sunday evening. There were no others and we talked for long of everything ranging from politics to atoms. He is the most charming man one can think of, proud and brilliant, and at the same time mild and modest … I was more fascinated by him than I can express in words and I so much wanted that you could have seen him. I think I learn a lot by meeting so many and so diverse people.

Earlier the same day Bohr had a conversation with an Amherst professor concerning literature and poetry of which he wrote to Margrethe: "I talked about literature with a local Professor of English who told me many interesting things about the working of style and versification; he persuaded me to tell him everything I knew and had thought of concerning ancient Nordic poetry." The unnamed professor was not Frost but someone else, possibly George B. Churchill who taught English at Amherst until 1925. After Bohr had returned to Denmark, he wrote a letter to Olds thanking him and his wife for their hospitability: "It was a most delightful and interesting experience to which my thoughts will often return … Also the rest of my stay in

---

[26] Petersen 1963; Heisenberg 1971, p. 40. On Bohr and language, see Favrholdt 1993.

[27] Niels Bohr to Margrethe Bohr, October 24, 1923 (NBA).

America was a great pleasure to me although it was quite a strenuous time."[28] Bohr might have mentioned his meeting with Frost, but he did not.

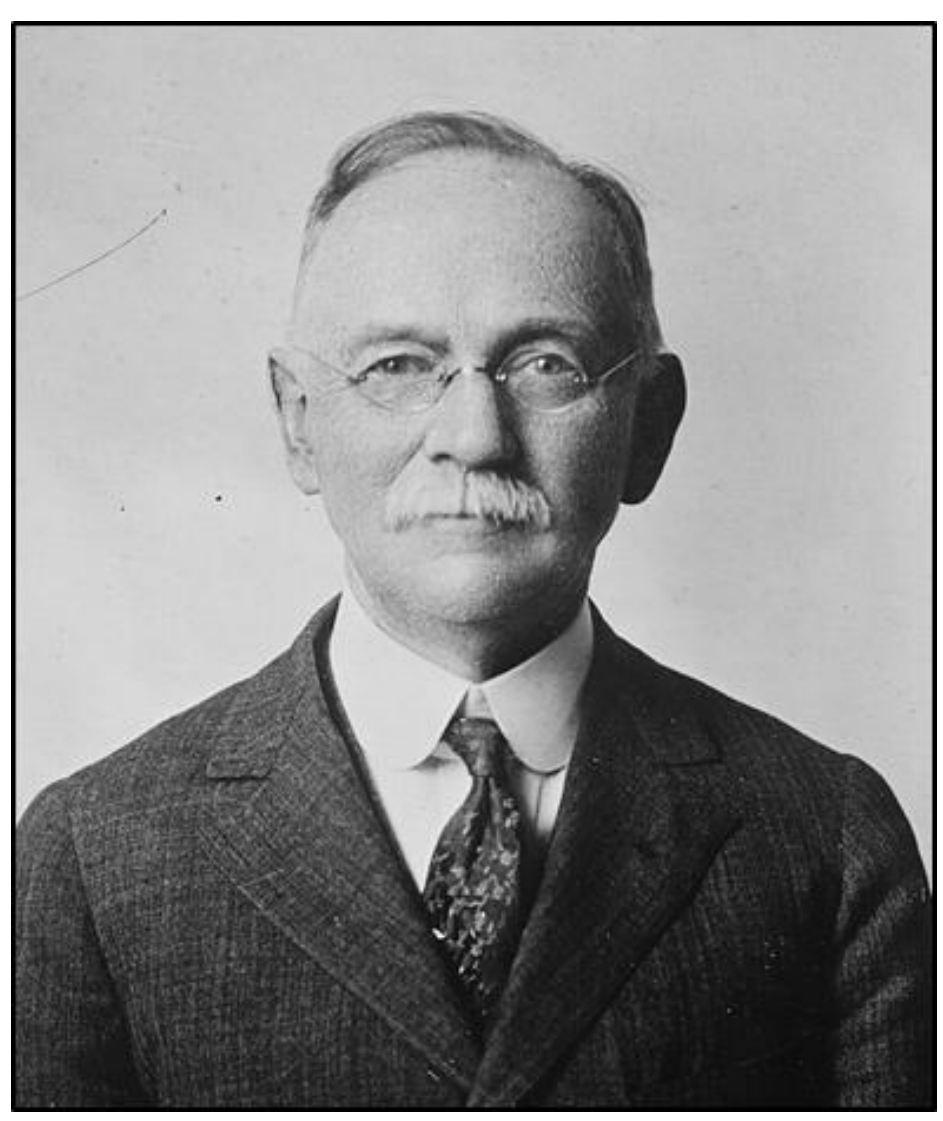

Fig. 5. George Daniels Olds (1853-1931).

There is one more relevant source which has escaped literary scholars. On February 24, 1928 Olds gave an address to the National Council of Teachers of Mathematics in which he discussed the nature and cultural value of mathematics. He recalled: "Some years ago Amherst College was host to Neils [*sic*] Bohr, of Copenhagen, at the time the most distinguished mathematical physicist in the world. He had said the latest and, some enthusiasts affirm, the last word on the structure of the atom." Olds (1928) continued:

> Few of us were able to understand his theory, but everyone caught the contagion of his presence. Like all communities, a college community finds the sudden appearance of a man of genius a disturbing influence. Bohr had not been with us many days before he expressed a strong desire to meet Robert Frost, a highly prized possession of the college. One Sunday evening I succeeded in getting the two men together at my fireside. There were a few awkward moments, some oppressive silence, then the talk began. It was ended only by my wife's opportune appearance to invite the two men into the dining room. They had been in touch with each other in a trice, the man of scientific imagination and the man of poetical imagination.

At the end of his address, Olds justified the decision to exposing ill-prepared college students to the thoughts of a Nobel laureate: "In the case of Bohr we could not follow him as he moved amid the intricacies of his theory, but his enthusiasm and evident genius seized us all with a grip that we shall never forget. For the sake of our pupils we must live from time to time on Parnassus and dwell with our muse."

---

[28] Bohr to Olds, December 31, 1923 (NBA).

## 7. The poet and the physicist: A casual connection

One may get the impression from Olds' account that Bohr was aware of Frost at an early date, perhaps before he arrived in Amherst. However, this seems to disagree with Bohr's letter to Margrethe casually mentioning "a Professor Frost here from Amherst." I suspect that he only expressed his "strong desire" to meet Frost after the dinner party on October 14, where Frost attracted his attention. As the poet would remember his one and only conversation with Bohr, so it made an impression on the Danish atomic physicist. At least, that is what his travel companion Hoyt recalled when looking back on the stay in Amherst almost forty years earlier:

> I do remember that he was very much impressed by Robert Frost who was at Amherst at that time, and there were several discussions with him. Unfortunately, I can't remember very well the content. But this apparently was something that Bohr always remembered, because whenever I would see him later on he would always bring up the name of Robert Frost, until the very last. This was something that was always on his mind.[29]

Bohr never mentioned Frost again in either his writings or correspondence and nor is there any indication that Bohr read some of Frost's poetry. He was clearly charmed by the poet's personality, even captivated by it, but that seems to have been it. After all, Frost's sceptical-ironic view about the nature and epistemic possibilities of physics stood in most respects in contrast to Bohr's.

In his letters to Margrethe, Bohr lyrically described the beautiful nature in New England and elsewhere, something which he much appreciated. On the other hand, he generally found the unfamiliar American way of life disturbing. Back in Copenhagen, he wrote to Rutherford that his visit to America had been "a very refreshing experience which gave me occasion to many thoughts." He then added: "Although one cannot avoid feeling how great the possibilities of the future are, I do not think I should like to live there all my life and miss the traditions which … give the colour to the life in the old countries."[30]

Frost may have been inspired by his 1923 conversation with Bohr, but if so he did not transform the inspiration into poetry. Nor did he ever address Bohr by letter or in any other way to learn more about his thoughts. Although he made poetical use of elements of quantum physics in several of his later poems and essays, in none of them did Bohr's atomic theory appear recognizably. The only possible exception is a couplet called "The Secret Sits" from 1942 (Frost 1942, p. 71):

> We dance round in a ring and suppose
> But the Secret sits in the middle and knows.

---

[29] Hoyt, American Institute of Physics interview, https://repository.aip.org/node/5540. While Hoyt referred to "several discussions," only the single one on 15 October is documented.

[30] Bohr to Rutherford, January 9, 1924, in Stolzenburg 1984, p. 487.

It takes some imagination to see in the two lines an allusion to the Rutherford-Bohr atomic model, but by interpreting 'we' as electrons, 'ring' as an orbit, and 'middle' as an atomic nucleus, this is what John Coletta and David Tamres have proposed and what is supported by a few other writers (Coletta and Tamres 1992). On the top of that, Barry Ahearn has suggested that parts of a poem Frost published in October 1923 "uncannily reflect a detail in Frost's encounter with Bohr" (Ahearn 2020, pp. 164-165). The poem in question is "I Will Sing You One-O," which includes the lines

> They left the storm
> That struck *en masse*
> My window glass
> Like a beaded fur.

Ahearn finds it "likely that Frost saw these lines in his poem as a version of what Bohr confirmed in their conversation." However, Frost's poem was published in the *Yale Review* on October 1, 1923, that is, before his conversation with Bohr. Although Ahearn's suggestion cannot be excluded, it seems far-fetched.

## 8. Conclusion

By charting Bohr's travels, lectures and meetings during his stay in North America in the autumn of 1923, one gets a vivid impression of how quantum atomic theory prior to quantum mechanics was received outside the narrow circle of scientific experts. Bohr's correspondence with his wife Margrethe is in this and other respects of considerable interest. A substantial part of the paper focuses on the transient connection between a great physicist, Niels Bohr, and a great poet, Robert Frost, as it occurred at Amherst College in October 1923. The new knowledge concerning their meeting adds an aspect not only to the history of physics but also to the history of literature.

**Acknowledgments**. I want to thank Vilhelm Bohr for granting me permission to use classified material from the Niels Bohr Archive, and Rob Sunderland for his help with respect to this and other archival materials.